\documentclass[twocolumn,aps,prl,superscriptaddress,floatfix,nofootinbib]{revtex4-1}
\usepackage{amsmath}
\usepackage{amsfonts}
\usepackage{amssymb}
\usepackage{graphicx}
\usepackage{color}
\usepackage{hyperref}
\usepackage{bm}

\begin{document}

\title{Statistical Majorana Bound State Spectroscopy}

\author{Alexander Ziesen}
\affiliation{JARA Institute for Quantum Information, RWTH Aachen University, 52056 Aachen, Germany}

\author{Alexander Altland}
\affiliation{Institut f\"ur Theoretische Physik, Universit\"at zu K\"oln, Z\"ulpicher Stra\ss e 77, 50937 K\"oln, Germany}
\author{Reinhold Egger}
\affiliation{Institut f\"ur Theoretische Physik,
Heinrich-Heine-Universit\"at, D-40225  D\"usseldorf, Germany}

\author{Fabian Hassler}
\affiliation{JARA Institute for Quantum Information, RWTH Aachen University, 52056 Aachen, Germany}

\date{\today}

\begin{abstract}
Tunnel spectroscopy data for the detection of Majorana bound states (MBS) is
often criticized for its proneness to misinterpretation of genuine MBS with
low-lying Andreev bound states. Here, we suggest a protocol  removing this
ambiguity   by extending single shot measurements to sequences performed at
varying system parameters. We  demonstrate how such sampling, which we argue
requires only moderate effort for current experimental platforms, resolves the
statistics of Andreev side lobes,  thus providing compelling evidence for the
presence or absence of a  Majorana center peak. 
\end{abstract}
\maketitle

{\it Introduction.---}About a decade after the first proposals for MBS
 engineering in topological quantum devices
 \cite{Alicea2012,Leijnse2012,Beenakker2013,DasSarma2015}, numerous reports of
 experimental signatures have been published, see, e.g.,
 Refs.~\cite{Mourik2012,NadjPerge2014,Lutchyn2018,Liu2018,Li2022,StationQ2022}.
 However, opinions remain divided as to whether ``Majoranas have been seen" or
 not. Broadly speaking, experiments aimed at MBS detection can be categorized
 into two groups: tunnel spectroscopy detecting midgap resonances caused by the
 assumed presence of an MBS
 \cite{Sengupta2001,Law2009,Flensberg2010,Sun2016,Zazunov2016}, and experiments going
 after unambiguous intrinsic properties of topological states, from
 unconventional noise correlations \cite{Demler2007,Nilsson2008,Golub2011,Haim2015,Liu2015,Tripathi2016,Jonckheere2017,Jonckheere2019,Manousakis2020}
 to full-feathered braiding protocols
 \cite{Aasen2016,Beenakker2020,Flensberg2021,Sbierski2022}. While the second
 group remains at the level of theoretical proposals, the former are
 straightforwardly realizable as a part of the core  MBS experiments. However,
 the downside is that tunnel spectroscopy data can be prone to
 misinterpretation. Among various other candidates for midgap signatures, pairs
 of conventional Andreev bound states --- which in symmetry class D
 \cite{Altland1997,Beenakker2015} superconductor  environments\footnote{
Strictly speaking, the system is either in class B or in class D depending on whether a MBS is present or not. For simplicity, we refer to both cases as ``class D''.} have a tendency to
 cluster around zero energy --- may leave experimental signatures hard to
 distinguish from a single MBS \cite{Bagrets2012,Liu2012,Aguado2017,Moore2018,Vuik2019,Prada2020,Valentini2021,Yu2021}.
 At any rate, as witnessed by the current debate on the ``topological gap
 protocol'' by the Microsoft Quantum team
 \cite{StationQ2022,Frolov2022,Akhmerov2022}, the community at large does not
 appear to be ready to take tunnel spectroscopy signatures, even of high
 quality, as unambiguous evidence for MBS formation.

In this paper,  we propose a relatively straightforward upgrade from single shot
tunnel spectroscopy measurements to parametric sequences of measurements. Their
realization for individual samples neither requires essential new hardware nor
measurement protocols beyond what is already available. We argue that the
compounded measurement data collected by statistical tunnel spectroscopy
\textit{does} contain compelling evidence for or against MBS formation.
Crucially, both the presence and the absence of an MBS will leave unique
imprints, provided the required statistical resolution has been met. A second
key feature is that disorder or device imperfections, usually considered as
unwelcome obstructions to MBS observability
\cite{Akhmerov2011,Wimmer2011,Brouwer2011,Neven2013,Diez2014,Haim2019}, here assume the
role of a resource: our approach works best for significantly disordered
systems.  

To understand its principle, we need to recall a few signatures of the spectrum
of class D superconductors \cite{Altland1997,Beenakker2015,Bagrets2012}. In confined
geometries subject to disorder or other sources of ``integrability breaking",
the Andreev spectrum is discrete, symmetric around zero energy, and subject to
statistical level correlations. Specifically, in the absence of topological
midgap states, Andreev bound states exhibit a slight statistical tendency to
attraction to zero energy, while they repel amongst themselves. Conversely, if a
topological midgap state is present, Andreev states get pushed away from zero
energy, and still repel amongst themselves. These signatures find a quantitative
representation in the ensemble-averaged spectral density \cite{Bagrets2012,Beenakker2015},
\begin{equation}\label{eq:sym_classes}
    \langle \rho(\epsilon)\rangle = \frac{1+c}{2}\delta(\epsilon)+ \frac{1}{\delta_\epsilon} \left(1-c\, \frac{\sin(2\pi\epsilon/\delta_\epsilon)}{2\pi \epsilon/\delta_\epsilon}\right),
\end{equation} 
where $\delta_\epsilon$ is the average (Andreev bound state) energy-spacing and
$c=+1$ ($c=-1$) in the presence (absence) of a MBS. The sinusoidal oscillations
in Eq.~\eqref{eq:sym_classes} describe a  tendency of the spectrum to
``crystallize'' into a statistically uniform sequence around zero, with
diminishing ($\sim \epsilon^{-1}$) rigor.  
Equation \eqref{eq:sym_classes} encodes a nonlocal fragmentation of the Hilbert
space and is obtained under the idealizing assumption of an infinite ensemble
subject to disorder strong enough to couple a large number of levels (random
matrix limit \cite{MehtaBook,Beenakker2015}). 

In experimental reality, there is no mathematical ensemble, disorder may not be quite so strong, and the recorded tunnel conductance data contains wave function fluctuations next to spectral signatures. Further, the MBS peak, if present, will be broadened by spectroscopic resolutions, temperature, and possibly other forms of environmental coupling. However, as we are going to argue, and demonstrate by numerical simulations, even relatively small \textit{sequences of measurements} performed for an engineered ensemble of configurations at limited resolution can reveal the principal signature of the spectral data: a statistical oscillation of period $\delta_\epsilon$ with opposite sign, depending on the presence or absence of Majorana states. In other words, a positive sign signal assumes the role of a control measurement revealing sufficient resolution for what in the presence of a MBS must flip sign to become a negative sign sequence. These rigidity patterns are deeply non-perturbative signatures of the class D spectrum, which in the $c=1$ case do require a \textit{single} midgap state (\emph{aka} Majorana). On this basis, we reason that a 
``smoking gun" signature is at hand.  While our approach in principle applies to arbitrary Majorana platforms, 
we illustrate it below for the example of a proximitized topological insulator (TI) 
slab pierced by a vortex \cite{Fu2008,Hasan2010,Ioselevich2012,Akzyanov2014,Flicker2019,Ziesen2021,Sandler2022}, where Andreev states correspond to Caroli-de~Gennes-Matricon subgap states \cite{Caroli1964}. 
In addition, in the SM \cite{supp}, we comment on alternative implementations in iron-based superconductors \cite{Petrovic2010,Zhang2016,Wang2018,Jiang2019,Machida2019,Kong2019,Zhang2018b,Zhu2020,Li2022}, 
planar phase-controlled Josephson junctions \cite{Hell2017,Halperin2017,Fornieri2019,Ren2019}, or
semiconductor hybrid nanowires \cite{Lutchyn2018,Flensberg2021,Marra2022}.  

\begin{figure}[tb]
  \centering
\includegraphics[width = \linewidth]{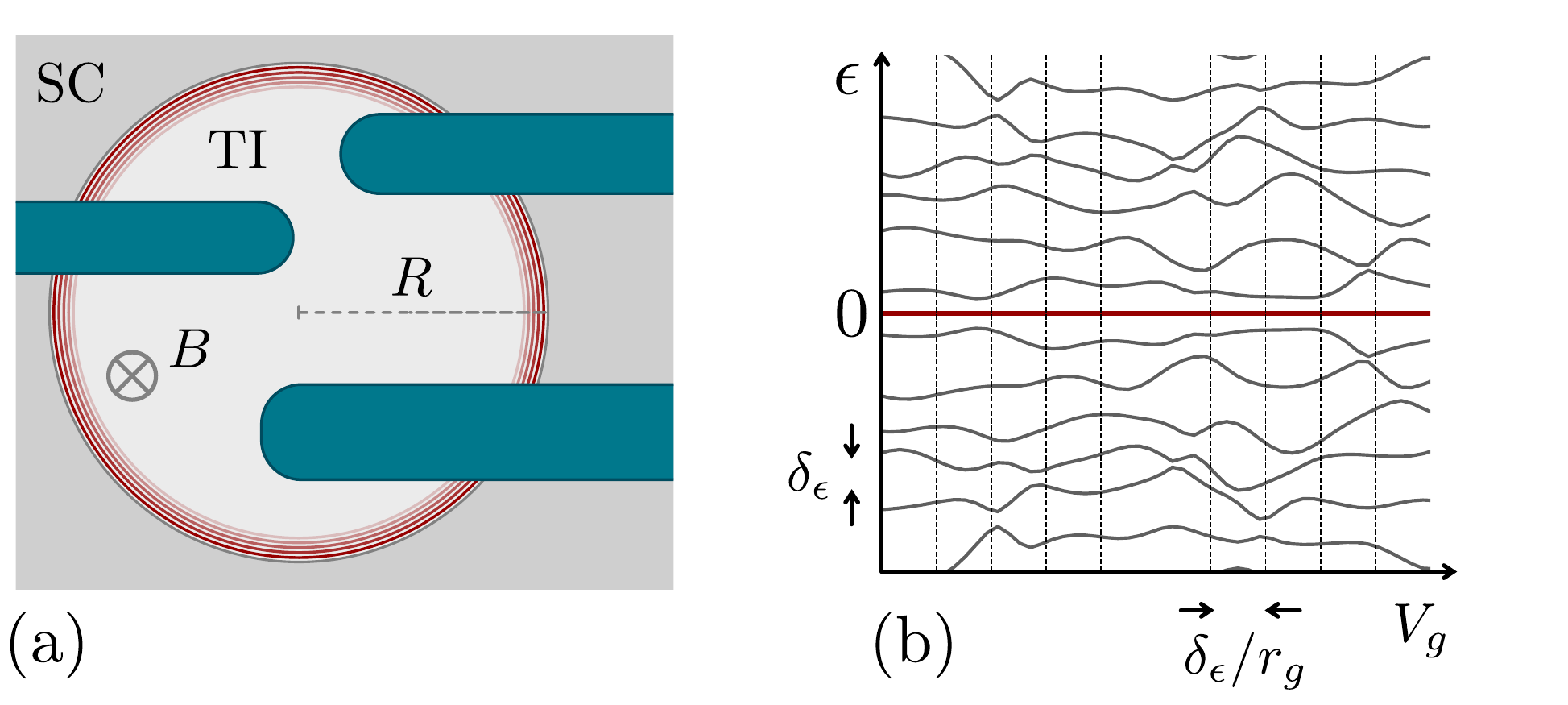}
\caption{Statistical spectroscopy setup. Left: A vortex is defined by a TI 
coated with an $s$-wave superconductor (SC) except for  a region of
radius $R$ which is threaded by $\nu$ magnetic flux quanta. Electrostatic finger
gates effectively change the disorder configuration. Red lines indicate spatial support of a Majorana edge mode. 
Right: In-gap spectrum vs gate voltage $V_g$ obtained by simulating the setup in (a). By varying $V_g$, Andreev state energy levels change on a scale set by the level spacing $\delta_\epsilon$. Sequences of independent disorder realizations are separated by $\delta V_g\approx \delta_\epsilon/r_g$ (see main text) as marked by the vertical lines. }
\label{fig1}
\end{figure}

\emph{Statistical spectroscopy principles.---}We propose a protocol where an
effectively averaged spectral density $\langle \rho(\epsilon)\rangle$ is
obtained by variation of external  control parameters. To understand the principle, we note that if 
integrability is broken by impurities and/or asymmetric system boundaries, the variation of any system parameter 
will result in new realizations of the chaotic scattering potential \cite{Goldberg1991}. 
Similar approaches have previously been applied in semiconductor devices \cite{Zumbuhl2002} and nanowires \cite{Contamin2022} for generating effective ensemble averages of
the tunneling conductance. As concrete example, we here formulate the approach for a TI vortex, cf.~Fig.~\ref{fig1}(a): 
An $s$-wave superconductor is deposited on a TI surface except for a circular region of radius $R$. Through this region an integer number, $\nu$, of superconducting flux quanta $\Phi_0 =
\pi/e$ is threaded ($\hbar=1$ throughout). For odd parity of $\nu$,  this  synthetic vortex binds a zero energy MBS \cite{Hasan2010}.

Variations in the voltage of nearby finger gates, $V_g$, parametrically change
the  system Hamiltonian. Even in the absence of ``intrinsic'' disorder, they break integrability and realize an effective ensemble
average, provided the perturbation is strong enough to effectively scramble the
spectrum of vortex states, cf.~Fig.\ref{fig1}(b). To estimate the required
voltage variations, we make the conservative assumption that the Coulomb
interaction across the vortex is strongly screened, and that only local wave
functions right under the geometric finger gate surface are susceptible to the
perturbation. To first order in perturbation theory, this leads to the estimate
$\left\langle \Psi |\delta V_g|\Psi \right\rangle \approx \delta V_g \int_g d^2r
|\Psi_\mathbf{r}|^{2}\approx r_g \delta V_g$ for the distortion of the energy, $\epsilon$, of individual states.
Here, the integral extends over the area underneath the gate, we assume approximate statistical uniformity of the
wave function modulus, and \(0\le r_g\le 1 \) is the fraction of the  gate area
relative to that of the vortex. Variations \(\delta V_g\gtrsim \delta_\epsilon/r_g \) 
strong enough that the perturbation exceeds the level spacing, $\delta_\epsilon$, 
effectively define a new realization of the spectrum, cf.~Fig.\ref{fig1}(b).  

Provided the broadening, \(\kappa\), of Andreev states due to disorder exceeds
the level spacing \(\delta_\epsilon\), we expect level repulsion, and in the
consequence the emergence of the spectral density in Eq.~\eqref{eq:sym_classes} upon
averaging over an ensemble. Presently, this ensemble average is realized by
sampling a large number of configurations distinguished by changes
\(\delta V_gr_g /\delta_\epsilon=\mathcal{O}(1)\), and subsequently collecting
the results in a histogram. 

In a concrete experiment where each level spacing is divided into \(N_b\) bins and 
the number of runs is $N_r$, an average number of $n_b=N_r/N_b$ levels will be counted per bin. 
This number is subject to statistical fluctuations \(\delta n_b\sim\mathcal{O}(n_b^{1/2})\). 
To obtain a reliable result, the relative fluctuation \(\delta n_b/n_b\) must be smaller than the relative change
\(|\rho(\epsilon)-\rho(\infty)|/\rho(\infty)\), computed according to Eq.~\eqref{eq:sym_classes}. 
A straightforward estimate for, say, \(\epsilon\approx 2\delta_\epsilon\) leads to the conclusion that $N_r \sim 10^2 N_b$ runs are required to obtain statistical certainty.  

\begin{figure}[tb]
  \centering
\includegraphics[width = \linewidth]{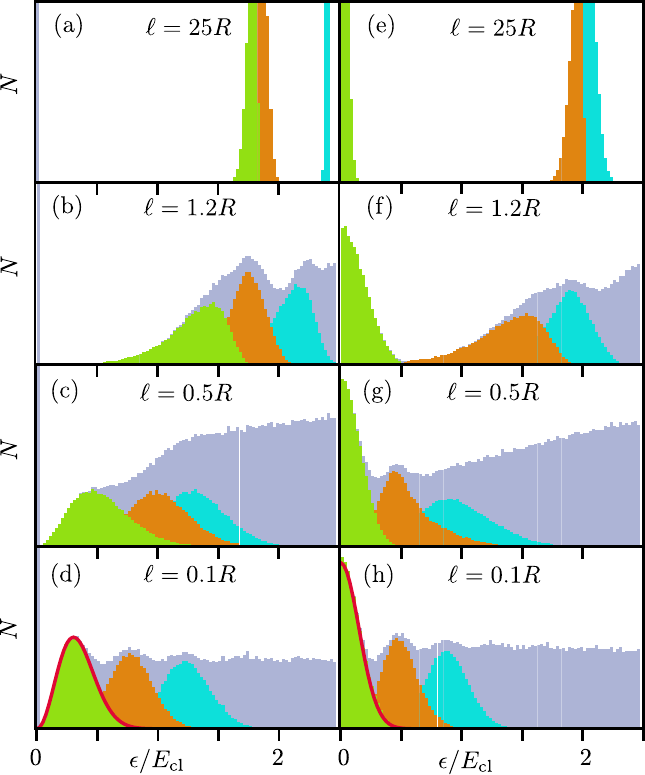}
\caption{Histogram for the positive energy levels of the TI vortex, with energies in units
of $E_{\rm cl}=v/R$. Panels (a)--(d) [(e)--(h)] show numerical results in the presence [absence] of a MBS with 
increasing disorder strength as obtained by diagonalizing $H_{\rm BdG}$ for $N_r=5\times 10^4$ disorder realizations.  
Green, orange and cyan colors refer to the three lowest  levels, all others are represented by the grey background. 
For weak disorder, $\ell \gg R$ [panels (a,e)], the averaged spectral peaks lie isolated,
they begin to overlap when  $\ell\sim R$ [panels (b,c) and (f,g)], and finally
combine to a continuum described by Eq.~\eqref{eq:sym_classes} at \(\ell
< R\) [panels (d,h)]. The statistics of the lowest level is accurately described
by the spacing distribution $P(\epsilon)$ (red curves) discussed in the text. }
\label{fig2}
\end{figure}

\emph{TI vortex.---}In the following, we test the statistical protocol for the TI vortex setup in Fig.~\ref{fig1}(a). 
The single-particle Bogoliubov-de~Gennes Hamiltonian describing the proximitized TI surface is given by  
\cite{Hasan2010}
\begin{equation}\label{eq:hamiltonian}
    H_{\mathrm{BdG}} = ( v \bm{p}\cdot \bm{\sigma}-\mu)\tau_z + \mathop{\rm Re}\Delta(\bm{r})\,\tau_x - \mathop{\rm Im}\Delta(\bm{r})\,\tau_y,
\end{equation}
where  $v$ is the surface-state velocity, \(\mu\) the chemical potential, and
Pauli matrices $\tau_i$ ($\sigma_i$) act in particle-hole (spin) space. In the
London gauge, the pair potential is $\Delta(\bm{r}) = |\Delta(r)| e^{-i\nu
\theta}$, with polar coordinates $(r,\theta)$ relative to the vortex
center and $\Delta(r)=\Delta\, \Theta(r-R)$ assumed as step function like. The
Hamiltonian \eqref{eq:hamiltonian} satisfies particle-hole symmetry,
\(\mathcal{C}H_{\mathrm{BdG}}\mathcal{C}^{-1}=-H_{\mathrm{BdG}}\) with
\(\mathcal{C}=\sigma_y \tau_yK\) and $K$ complex conjugation, placing it into
symmetry class D. (For completeness, we mention that in the field free case,
\(\nu=0\), we also have time reversal symmetry, \(\mathcal{T}H_{\mathrm{BdG}}\mathcal{T}^{-1}=H_{\mathrm{BdG}}\) with
\(\mathcal{T}=i\sigma_yK\), implying an upgrade to class DIII.)

We add disorder to the vortex core $r<R$ in the form of a  
 Gaussian correlated random potential $V(\bm r) \tau_z$ with zero mean and
 variance $\langle V(\bm{r}) V(\bm{r}') \rangle = \gamma^2
 \delta(\bm{r}-\bm{r}')$. The corresponding scattering mean free path computed
 in Born approximation is given by $\ell = v/\kappa=(v/\gamma)^2 R$ \cite{supp}. 
 Comparison with the bound-state spacing of the clean vortex,
 $\delta_\epsilon\simeq v/R\equiv E_\mathrm{cl}$, shows that the threshold to
 strong disorder mixing, \(\delta_\epsilon\sim \kappa\), is reached when \(\ell
 \sim R\), i.e., when the quasi-particle motion crosses over from ballistic to
 diffusive. For stronger disorder, the characteristic level spacing shrinks to
 \(\delta_\epsilon \simeq (\ell/R)(v/R)\equiv E_\mathrm{Th}\), i.e., from the
 inverse of the ballistic time of flight, $E_\mathrm{cl}$, to the  inverse of the
 diffusion time across the vortex, \(E_\mathrm{Th}\), see the SM \cite{supp} for details. 

Our statistical approach to MBS spectroscopy works for disorder beyond the ballistic/diffusive
 threshold. To illustrate this point, Fig.~\ref{fig2} shows data histograms obtained from
 $N_r = 5\times 10^4$ disorder realizations and for disorder strengths ranging
 from an almost perfectly ballistic regime, \(\ell=25 R\), to a diffusive one with
 \(\ell=0.1 R\). The columns on the left (right) are for a vortex with (without)
 MBS, realized here by setting $\nu=1$ ($\nu=2$). 
 In the ballistic regime, we observe weakly broadened states with
 spacings varying strongly at scales \(\sim E_\mathrm{cl}\). Upon crossing into
 the diffusive regime, they start to overlap, along with a tendency towards a
 more uniform spacing --- the level crystallization symptomatic for quantum
 chaotic spectra.  
 
 Real experiments have access to the cumulative contribution of all levels, here
 indicated in grey, where we observe the gradual approach to the profile in Eq.~\eqref{eq:sym_classes}, as well as to the distribution of individual levels, cf.~the green/orange/cyan histograms for the lowest three positive energy levels. 
 For disorder deep in the diffusive regime, we expect the statistics of these levels 
 to be described by the principles of random matrix theory \cite{MehtaBook,Beenakker2015}. 
 Specifically, for class D one expects the probability distribution for 
 the lowest lying level in the case with [without] MBS
 to be given by $P(\epsilon) \propto \epsilon^2 \exp(-\epsilon/2)$ 
 $[P(\epsilon)\propto \exp(-\epsilon^2/2)]$ \cite{Altland1997,Beenakker2015,supp}. Figure \ref{fig2} shows that
 these distributions, indicated as red curves, are clearly realized by the
 disordered vortex in the strong disorder regime. However, the most important
 conclusion is that the presence or absence of a MBS is clearly resolved via
 the statistics of the cumulative histogram, provided the focus of attention is shifted
 to the side bands, and the disorder is sufficient to induce inter-level
 correlations.

\emph{Experimental reality.---}The analysis above assumed arbitrary energy resolution, and averaging over a 
large number \(\mathcal{O}(10^5)\) of realizations. What happens under less
ideal conditions?    
In an experiment, the potential $V({\bm r})$ describing impurities or scattering
off device irregularities is fixed and different realizations of the spectrum are
generated by variation of externally adjustable parameters. In the TI vortex,
the magnetic field strength is likewise fixed, which leaves gate electrodes as the next
best choice for generating a parameter set. To generate \(N_r
\sim 10^2 N_b\) samples required for \(N_b\) bins per level
spacing (see above estimate),
one may need to work with \(f\) finger gates and the resulting \(f\)-dimensional
parameter space. As the gate voltages are meant to mimic ``disorder'', it is best to 
use an asymmetric geometric design as indicated in Fig.~\ref{fig1}. Electrodes
with large  electrode-to-vortex area ratio, \(r_g\), will generate
optimal sensitivity of energy levels, \(\sim r_g V_g\). 

 The simulations discussed in the following were performed for
\(N_b=10\) bins, requiring \(N_r=\mathcal{O}(10^3)\) runs. We worked with
\(f=3\) electrodes \cite{supp}, and varying each of their voltages over a range \(\delta V_g
\sim \delta_\epsilon/r_g \) generated the parameter space for up to \(N_r\approx
10^4\) statistically independent samples~\footnote{If no independent information on  the characteristic level spacing
\(\delta_\epsilon\) is available, the latter may be estimated by measuring the spacing
between peaks in the averaged spectral density.}. Finally, we  account for the
broadening of individual levels due to  temperature or environmental coupling by introducing
a Lorentzian line width, \(\delta(\epsilon)\mapsto (\Gamma/\pi)/(\epsilon^2 +
\Gamma^2)\), with \(\Gamma=0.05E_\mathrm{cl}\). Here, \(\Gamma\lesssim
\delta_\epsilon\) is required to resolve the oscillatory pattern of the target
spectral density, i.e., our method requires the resolvability of individual
states~\footnote{Another source of uncertainty is due to the fact that tunnel
spectroscopy measures point conductance, i.e., a quantity proportional to the
product of spectral density and wave function moduli, where the latter remain
unknown. However, one may expect that in a  large data set, these variations
efficiently average out.}. 

\begin{figure}[tb]
  \centering
  \includegraphics[width = \linewidth]{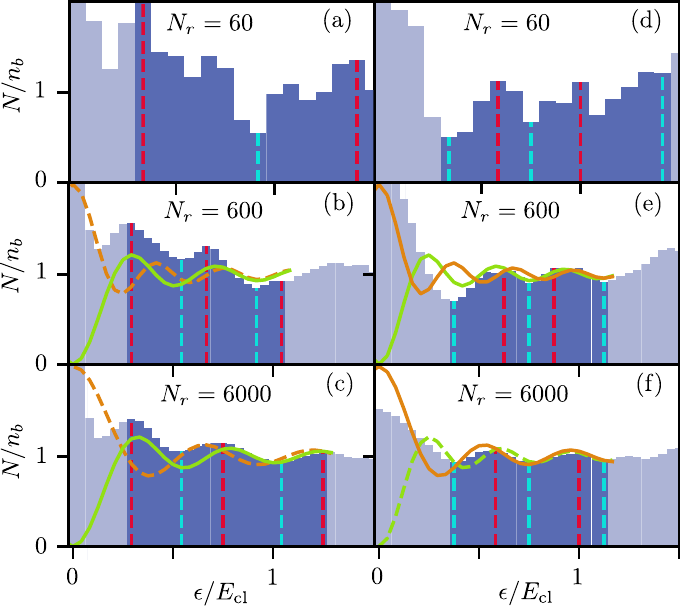}
  \caption{Histogram as in Fig.~\ref{fig2} but for a single realization of $V({\bm
  r})$ with $\ell=0.1R$, using an increasing number of $N_r$ samples with (left column)
  and without (right) MBS. The first five extrema are indicated
  by dashed vertical lines. The green (orange) curves are fits to
  Eq.~\eqref{eq:sym_classes} for \(c=1\) (\(c=-1\)) within the dark blue regions, i.e., with the \(\delta\)-peak removed,  using  \(\delta_\epsilon\) as single fit parameter.  
  Inferior fits are shown as dashed curves.}
  \label{fig3}
  \end{figure}

Given this setup, the minimal goal is a statistically sound distinction between
the cases \(c=\pm 1\) in Eq.~\eqref{eq:sym_classes}. Figure~\ref{fig3} illustrates how  the two cases
are distinguished through a phase shift in the  oscillatory spectral
density \textit{at finite energy}. In either case, a midgap peak is present
(caused by a broadened MBS for \(c=+1\), or a statistical accumulation of
Andreev states for \(c=-1\)). While these two peaks are difficult to distinguish, our
method focuses on the spectrum away from the center. We  note that in either
case, the average spectral density contains a sequence of extrema at
\(\epsilon\sim  \frac{2n+1}{4}\delta_\epsilon\). The difference is that this sequence starts with a maximum (for \(c=1\)) or a minimum (for \(c=-1\)). A more refined signature is obtained by subtracting a constant
background and fitting the remaining oscillatory signal for
the first, say, five extrema to Eq.~\eqref{eq:sym_classes}, using \(\delta_\epsilon\) as 
single fit parameter. 

Fig.~\ref{fig3} shows data processed in this way for increasing number of runs \(N_r\), 
either with (left column) and without (right) MBS. The quality of
the data may be assessed, e.g., by calculating the sum of squared distances between  
the extrema of the fit function and the data.  For too low sample number, e.g., 
for $N_r=60$, no unambiguous pattern of extrema is identifiable. At $N_r=600$
samples, side lobes begin to emerge, but a reliable assignment of extrema is still difficult to ascertain. 
However, for $N_r=6000$, the extremal energies are evenly spaced,
and the squared distance fit accurately determines the correct sign of $c$.
Additional information on system parameters, such as knowledge of the effective broadening $\Gamma$, may
be exploited to develop more informed fitting protocols for the ensemble averaged data. 
However, we found that such refinements lead only to minor improvements of the results.

Let us briefly comment on the experimental feasibility of the TI vortex setup.
Generally speaking the vortex area should
be chosen small enough that its quantized levels can be resolved, and large
enough that neighboring levels are coupled by disorder and gate variations. With 
$\delta_\epsilon\approx E_\text{Th} =v \ell /R^2$, and given typical
values $v\approx 5\times 10^{5}\,$m/s,  $\ell\approx 20$~nm 
\cite{Ando2011}, with spectral resolution $\Gamma\approx 30\,\mu$eV, 
one needs to have $R\alt 3\,\mu$m. Choosing $R=300\,$nm and finger gates
of width $\approx 50\,$nm, we have $\ell/R \approx 0.1$. For these values,
individual levels can be distinguished and a few finger gate electrodes could be placed
over the vortex core. We are thus confident that the requirements for our
proposal to work are met by existing setups. 

\emph{Conclusions.---}We have proposed a novel scheme for the detection of MBS
in existing device structures which combines
tunnel spectroscopy with elements of statistics. The focus of attention
is here shifted from the center peak ubiquitous in spectroscopic data --- which is
notorious for its misinterpretability --- to the pattern of side bands. The
unavoidable presence of effective disorder becomes a resource in that it induces
correlations between levels which, upon averaging over different parametric
realizations, lead to the effectively crystalline structure in 
Eq.~\eqref{eq:sym_classes}. The latter originates in a combination of statistics
and topology which is unambiguously linked to the presence or absence of a
MBS, even if the latter cannot be clearly identified in isolation. Another
advantage of the approach is that it includes its own validation: If neither
the positive, \(c=1\), nor the negative, \(c=-1\), signal can be resolved, the
method has not been implemented with sufficient accuracy. 
The principal conditions for it to work are resolvability of individual levels (where
one may argue that this condition must be met anyway for the  MBS to become a
useful resource), sufficient statistics provided by at least
\(\mathcal{O}(10^3)\) runs, and effective disorder strong enough to cause level
correlation. (If the ``native'' disorder is too weak, one may contemplate
lowering the level spacing by increasing the vortex size for diagnostic
purposes.) These criteria are realistic for the vortex platform, and we are
confident that the same holds for other realizations, such as planar Josephson
junctions, leaving sufficient freedom for the placement of gate electrodes.
We conclude that this approach has the potential to settle the issue of MBS existence
with available measurement protocols and hardware.  

\begin{acknowledgments}
We thank Y. Ando, C. Marcus, J. Schluck, and S. Vaitiekenas for discussions.
We acknowledge funding by the Deutsche Forschungsgemeinschaft (DFG, German Research Foundation),
Projektnummer 277101999 -- TRR 183 (AA and RE, projects A01, A03, B04, and C01), under project No.~EG 96/13-1,
and under Germany's Excellence Strategy -- Cluster of Excellence Matter and Light for 
Quantum Computing (ML4Q) EXC 2004/1 -- 390534769.
\end{acknowledgments}

\bibliographystyle{aipnum4-1}
\bibliography{biblio}

\end{document}


\title{Supplementary Material to ``Statistical Majorana Bound State Spectroscopy''}

\author{Alexander Ziesen}
\affiliation{JARA Institute for Quantum Information, RWTH Aachen University, 52056 Aachen, Germany}

\author{Alexander Altland}
\affiliation{Institut f\"ur Theoretische Physik, Universit\"at zu K\"oln, Z\"ulpicher Stra\ss e 77, 50937 K\"oln, Germany}
\author{Reinhold Egger}
\affiliation{Institut f\"ur Theoretische Physik,
Heinrich-Heine-Universit\"at, D-40225  D\"usseldorf, Germany}

\author{Fabian Hassler}
\affiliation{JARA Institute for Quantum Information, RWTH Aachen University, 52056 Aachen, Germany}

\begin{abstract}
We here derive the distribution function $P(\epsilon)$ from random matrix theory (see Sec.~\ref{sec1}),  
discuss the mean free path $\ell$ (see Sec.~\ref{sec2}), 
provide details on our numerical calculations in Sec.~\ref{sec3},
and comment on the application of our approach to alternative Majorana platforms in Sec.~\ref{sec4}.
Equation or figure numbers containing the index 'M' refer to the main text.
\end{abstract}
\maketitle

\section{Distribution functions}\label{sec1}

Analytical results for the distribution function $P(\epsilon)$ 
of the lowest finite-energy level $\epsilon$  
can be obtained by random matrix theory arguments \cite{MehtaBook}.
In practice, it is sufficient to calculate $P(\epsilon)$ for
the smallest nontrivial matrix dimension $d$ for the corresponding
class D random matrix ensemble, where $d=3$ ($d=2$) with (without) a MBS.  
Using independent real-valued normal-distributed random variables, $\{a,b,c\}$, the respective 
random matrices are written as
\begin{equation}
 V_{d=3} = \begin{pmatrix} 0 & -a & -b \\ a & 0 & -c \\ b & c & 0 \end{pmatrix},   
 \quad V_{d=2} = \begin{pmatrix} 0 & -a \\ a & 0\end{pmatrix}.  
\end{equation}
The particle-hole symmetric spectrum is then given by 
$\lbrace 0,\pm\lambda_{3}=\pm \sqrt{a^2+b^2+c^2} \rbrace$ for $d=3$, and by 
$\lbrace \pm \lambda_{2}=\pm |a|\rbrace$ for $d=2$. 
Expressing the dependence on $\{a,b,c\}$ in terms of multivariate 
$\chi^2$-distributions, we obtain the distributions 
\begin{equation}
P(\lambda_{3}) = \sqrt{\frac{2}{\pi}}\lambda_{3}^2 e^{-\lambda_{3}^2/2},\quad
P(\lambda_{2}) = \sqrt{\frac{2}{\pi}} e^{-\lambda_{2}^2/2}.
\end{equation}
The energy distribution, $P(\epsilon)$, 
with $\epsilon= \lambda_{3}$ or $\epsilon =\lambda_{2}$, respectively, 
now follows as quoted in the main text.

\section{Mean free path}\label{sec2}

The mean free path $\ell$ connects the disorder strength $\gamma$ to experimentally accessible quantities 
for the TI vortex platform. Using the standard Born approximation
(which applies for our system at finite energy in the presence of
weak disorder \cite{Neven2013})
and the density of states 
$D_{2}(\epsilon)=|\epsilon|/(2 \pi  v^2)$ obtained from Eq.~(M2), 
the energy-dependent mean free path is given by
\begin{equation}\label{elldef}
    \ell = \frac{ v}{2 \pi \gamma^2 D_{2}(\epsilon)} = \frac{v^3}{\gamma^2 |\epsilon|}.
\end{equation}  
For low-energy in-gap states, Eq.~\eqref{elldef} is evaluated at $\epsilon=\delta_\epsilon$, with
$\delta_\epsilon$ the level spacing. Depending on the hierarchy of $\ell$ and $R$, 
$\delta_\epsilon$ is either given by the ballistic ($\delta_\epsilon\simeq E_{\rm cl}= v/R$) or 
by the diffusive ($\delta_\epsilon \simeq E_{\rm Th}= v\ell/R^2)$ Thouless scale.
In the respective limits, we then obtain
\begin{equation}
    \ell \approx \begin{cases} \frac{v^2}{\gamma^2} R, & \gamma \ll v , 
    \\ \frac{v}{\gamma} R, & \gamma \gg v.\end{cases} 
\end{equation}
Thus, $\delta_\epsilon$ and $\ell$ both are suppressed by a factor of $v/\gamma$ in the diffusive regime.

\section{Numerical simulation details}\label{sec3}

We present details about the numerical simulations employed to verify the gate-based sampling method. 
The full Bogoliubov-de~Gennes (BdG) Hamiltonian in the presence of the Gaussian random potential $V({\bm r})$
is  $H=H_{\rm BdG}+V({\bm r})\tau_z$ with $H_{\rm BdG}$ in Eq.~(M2). For an etched vortex, the radial profile of the pairing gap is well approximated by $\lvert \Delta(\bm{r})\rvert = \Delta\, \Theta(r-R)$. 
Moreover, for low-energy in-gap states, we can effectively capture the effects of the
superconductor as a boundary condition at $r=R$ which follows by sending $\Delta\to \infty$.
Below we use the quantum numbers $\alpha=(n,m)$, with the (integer or half-integer) angular momentum $m$ and 
the (integer) radial quantum number $n$. Introducing also the dimensionless radial variable
$x_\alpha = \epsilon_{\alpha}r/ v$ for finite energy $\epsilon_\alpha \neq 0$, the
corresponding eigenstates of $H_{\rm BdG}$ for $\mu=\Delta=0$ are expressed in terms of Bessel functions, 
\begin{equation}
    f_\alpha(\bm{r}) = \mathcal{N}_\alpha e^{i m \theta }\begin{pmatrix} 
    i e^{-i(\nu+1) \theta/2} J_{m-(\nu+1)/2}(x_{\alpha}) \\
    - e^{-i(\nu-1) \theta/2} J_{m-(\nu-1)/2}(x_{\alpha}) \\
   i c_\alpha e^{i(\nu-1) \theta/2} J_{m+(\nu-1)/2}(x_{\alpha}) \\
     c_\alpha e^{i(\nu+1) \theta/2} J_{m+(\nu+1)/2}(x_{\alpha})
     \end{pmatrix} ,
\end{equation}
where we use polar coordinates, the four-spinor convention $(\psi_{\uparrow},\psi_{\downarrow},\psi_\downarrow^\dagger,
-\psi_\uparrow^\dagger)$ in spin and particle-hole space, and a normalization factor ${\cal N}_\alpha$.
The superconducting boundary condition at $r=R$ leads to the conditions
\begin{eqnarray} \label{cond1}
 && \frac{   J_{m-(\nu-1)/2}(x_{\alpha})J_{m+(\nu+1)/2}(x_{\alpha})}{J_{m-(\nu+1)/2}(x_{\alpha})J_{m+(\nu-1)/2}(x_{\alpha})}=1 ,\\ \nonumber
 && c_\alpha = \frac{J_{m-(\nu+1)/2}(x_{\alpha})}{J_{m+(\nu+1)/2}(x_{\alpha})}  .
\end{eqnarray}
Once the energies $\epsilon_\alpha$ solving the first (implicit) equation in Eq.~\eqref{cond1} have been determined, 
the coefficients $c_\alpha$ follow from the second equation. 
In addition, the spectral problem admits $\nu$ zero-energy solutions, 
where an even number of such states can hybridize to form topologically trivial Andreev bound states
and one finds a single MBS for odd $\nu$.
In particular, for $\nu=1$, the zero-energy state (with $m=n=0$) is given by
\begin{equation}
    f^{(\nu=1)}_{0,0}(\bm{r}) =  \frac{1}{\sqrt{2}}\begin{pmatrix} 0 \\ -1 \\ i \\ 0
    \end{pmatrix},
\end{equation}
while for $\nu=2$, we find zero-energy states with angular momentum $m=\pm 1/2$, 
\begin{eqnarray}
    f^{(\nu=2)}_{0,-1/2}(\bm{r}) &=&
    \sqrt{\frac{2}{3 \pi}}
    \begin{pmatrix} 0 \\ e^{-i\theta}r/R \\ -i \\ 0
    \end{pmatrix} ,\\
    \nonumber
    f^{(\nu=2)}_{0,1/2}(\bm{r}) &=&\sqrt{\frac{2}{3 \pi}}
    \begin{pmatrix} 0 \\ i \\e^{i\theta}r/R \\ 0
    \end{pmatrix}  ,
\end{eqnarray}
which combine to form an Andreev bound state.

\begin{figure}[tb]
  \centering
\includegraphics[width=0.29\textwidth]{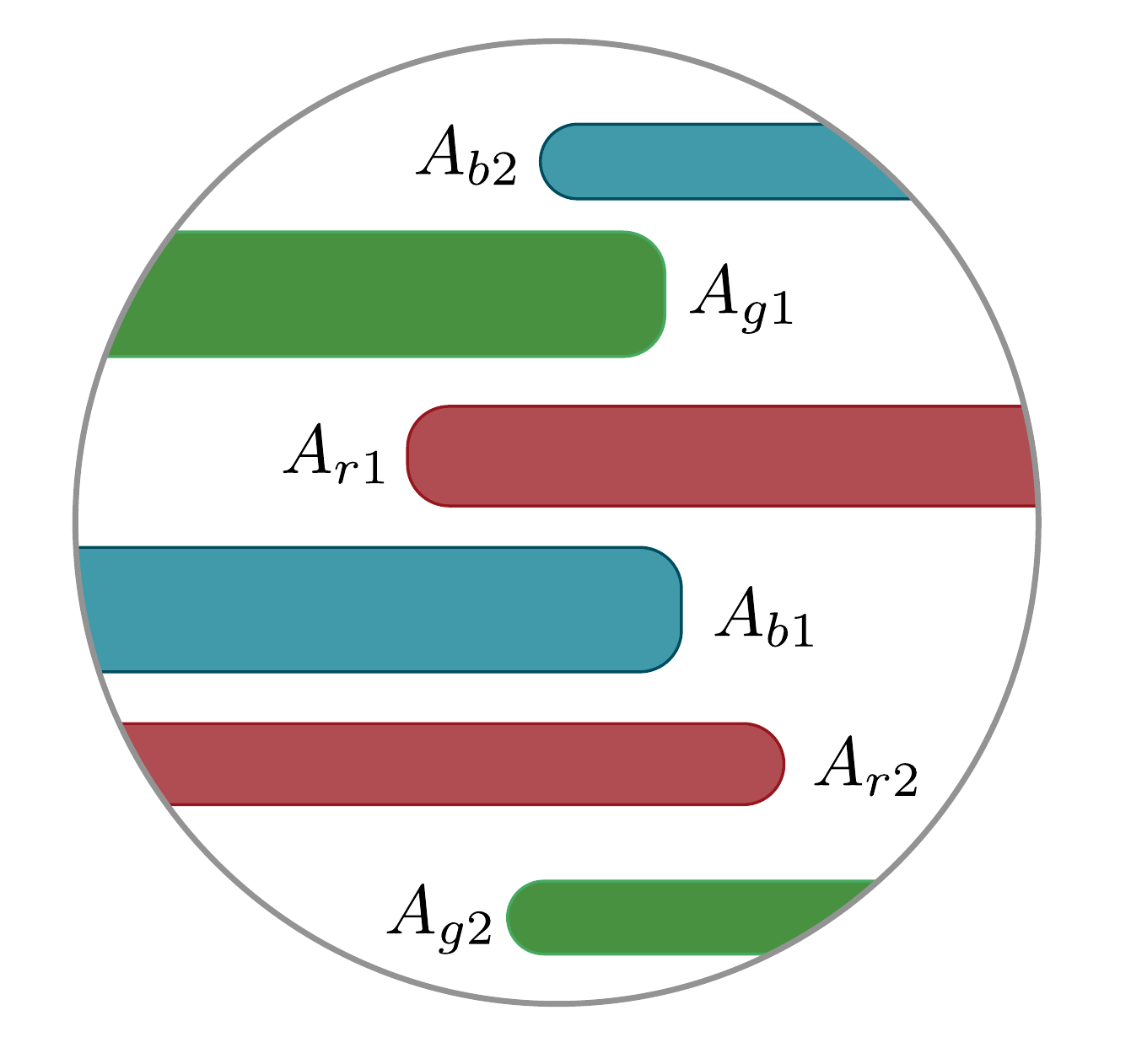}
\caption{TI vortex setup with six finger gates placed within the vortex core region of radius $R$. 
The gates cover areas of different sizes, $A_{cj}$, and are arranged in three pairs indicated by the colors $c\in \lbrace r,b,g \rbrace$. Gates forming a pair are referenced by $j \in \lbrace 1,2\rbrace$. }
\label{figS1}
\end{figure}

Next, we discuss the matrix elements of the disorder potential in the basis diagonalizing $H_{\rm BdG}$,
\begin{equation} \label{eq:disorder_2d_integral}
V_{\alpha\beta} = \int d\bm{r}\, f_\alpha^{\dagger}(\bm{r}) V(\bm{r})\tau_z f_\beta(\bm{r}) .
\end{equation}
In order to avoid the explicit evaluation of the slowly converging and computationally intensive 
spatial integrals in Eq.~\eqref{eq:disorder_2d_integral}, we use the fact that, since $V({\bm r})$ is Gaussian distributed, the matrix elements $V_{\alpha\beta}$ have 
only nontrivial second moments. However, these
matrix elements are correlated and therefore the full covariance matrices $\langle V_{\alpha \beta}V^{*}_{\gamma \delta}\rangle$
and $\langle V_{\alpha \beta}V_{\gamma \delta}\rangle$ are needed, where angular momentum conservation implies 
\begin{eqnarray}\label{eq:angular_disorder_separation}
 \langle V_{\alpha \beta}V^{*}_{\gamma \delta}\rangle &\propto &
 \delta_{m_\alpha-m_\beta, m_\gamma-m_\delta},\\ \nonumber
 \langle V_{\alpha \beta}V_{\gamma \delta}\rangle &\propto& 
 \delta_{m_\alpha-m_\beta, -(m_\gamma-m_\delta)} .
\end{eqnarray}
We note that matrix elements involving zero-energy states also satisfy Eq.~\eqref{eq:angular_disorder_separation}. 
With $\Delta m_{\alpha\beta} = m_\alpha-m_\beta$, we next observe that
$\langle V_{\alpha \beta}V^{*}_{\gamma \delta}\rangle\neq 0$ only if 
$\Delta m_{\alpha\beta}=\Delta m_{\gamma\delta}$. Similarly, $\langle V_{\alpha \beta}V_{\gamma \delta}\rangle\neq 0$ 
requires $\Delta m_{\alpha\beta}=-\Delta m_{\gamma\delta}$. 
Clearly, both covariances can be finite only for a disjoint set of disorder elements unless $\Delta m_{\alpha\beta}=\Delta m_{\gamma\delta} = 0$. 
It is then convenient to define 
\begin{equation}\label{eq:W_definition}
    W_{\alpha \beta} = \begin{cases} V_{\alpha \beta}, & \Delta m_{\alpha\beta}>0 , \\
    V_{\alpha \beta}^{*}, & \Delta m_{\alpha\beta}<0 , \\ 
    V_{\alpha \beta} + i(V'_{\alpha \beta})^{*}, & \Delta m_{\alpha\beta}=0 ,\end{cases}
\end{equation}
with statistically independent auxiliary variables $V'_{\alpha \beta}$ (not entering the Hamiltonian) which are distributed identically as $V_{\alpha \beta}$. 
The new variables $W_{\alpha\beta}$ fulfill $\langle W_{\alpha \beta}W_{\gamma \delta}\rangle = 0$, and therefore  
the problem has been reduced to a single Hermitian covariance matrix,
$C_{\alpha \beta \gamma\delta} = \langle W_{\alpha \beta}W^{*}_{\gamma \delta}\rangle$.
Using a singular value decomposition, 
$C= U \Sigma U^\dagger = L L^\dagger$ with $L = U \sqrt{\Sigma}$,
we can then map a set of \emph{independent} normal-distributed complex elements, 
$X_{\alpha \beta}$, onto correlated matrix elements, $W_{\alpha\beta} = (L X)_{\alpha \beta}$. 
By using the inverse mapping of Eq.~\eqref{eq:W_definition}, 
one obtains the matrix elements $V_{\alpha\beta}$ in a numerically efficient manner. 
We have verified that disorder matrices generated by this mapping obey the same distribution as the spatial integrals in Eq.~\eqref{eq:disorder_2d_integral}.

We finally describe the simulated gate setup used for generating the data 
in Fig.~M3, which is illustrated in Fig.~\ref{figS1} and 
contains six finger gates inside the vortex 
with area $A_{\rm vort}=\pi R^2$.  The gates are paired up into $f=3$ pairs, 
where each gate covers the area $A_{cj}$, with $c \in \lbrace r,b,g\rbrace$ indicating the ``color'' of a pair 
and $j \in \lbrace 1,2\rbrace$ numbering both gates forming the pair.
The latter are assumed to be oppositely charged by means of opposite gate voltage shifts, $\pm \delta V_c$. 
All finger gates should be spaced sufficiently far away from each other
such that direct tunneling processes between gate electrodes can be ruled out.
Discretizing the vortex area $A_{\rm vort}$ into a grid of infinitesimal area pieces 
$\delta A$, the perturbation strength due to a gate pair with voltage changes $\pm \delta V_c$ 
can then be estimated by summing over this spatial grid, where the gate voltage is constant for each finger gate and 
zero elsewhere.  We then arrive at
\begin{equation}\label{eq:comp_gate_based_pert}
\bra{\Psi}\delta V_c\ket{\Psi} \approx 
\frac{\delta V_c}{A_\mathrm{vort}}\left( A_{c1}  -A_{c2} \right) = r_c\delta V_c  .
\end{equation}
Equation~\eqref{eq:comp_gate_based_pert} 
 defines the area fraction, $r_g=r_c$, for this gate pair.
Independently tuning gate voltages on different gate pairs allows one to sample 
many disorder realizations, where an uncorrelated new sample is reached by
varying the respective gate voltage by $\delta V_c \approx \delta_\epsilon/r_c$.

\section{Alternative platforms}\label{sec4}

In the main text, we demonstrated our protocol on a numerically simulated proximitized TI vortex. 
We here discuss some of the most prominent alternative types of MBS hardware and speculate on their suitability for statistical spectroscopic analysis. 

\begin{figure}[tb]
  \centering
\includegraphics[width=0.47\textwidth]{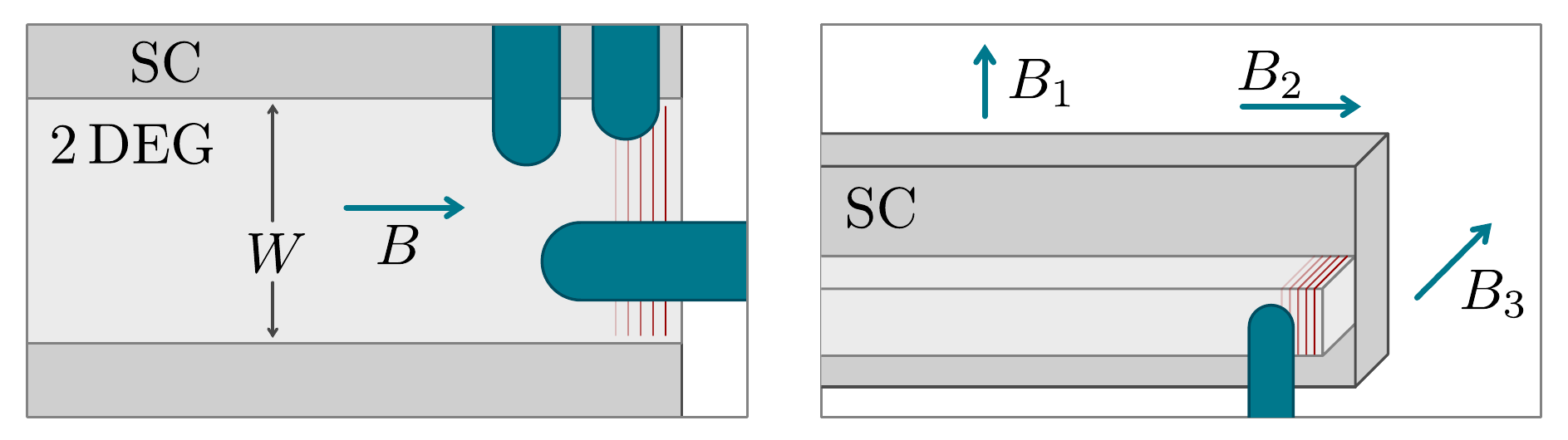}
\caption{Possible alternative realizations of statistical spectroscopy. Left: Planar Josephson junction. Right: Semiconductor quantum wire. In either case, a combination of electrostatic side gates and 
external magnetic fields may realize the ensemble parameter space. 
The fading red lines indicate the spatial support of a Majorana end state.}
\label{figS2}
\end{figure}

\emph{Planar Josephson junctions} [left panel in Fig.~\ref{figS2}] support 
low-energy bound states along the edge of a two-dimensional electron gas (2DEG) 
\cite{Hell2017,Halperin2017,Fornieri2019,Ren2019}.
The width, $W$, of the latter is adjustable  and may be 
increased to allow for the placement of a number of top gates, much as in the TI vortex case. At the same time, the level spacing is proportional to $1 /W$, meaning that a too wide junction will be in conflict with the required resolvability of levels. We thus have a tradeoff situation which calls for prior numerical optimization. 
However, generally speaking the planar Josephson junction appears to be a promising candidate for the realization of statistical spectroscopy. 

\emph{Proximitized nanowires with spin-orbit coupling} [right panel in Fig.~\ref{figS2}]
have been investigated for more than a decade as candidates for the realization
of Majorana end states, see Refs.~\cite{Lutchyn2018,Flensberg2021} for reviews. 
We see two principal factors relevant to the
applicability of statistical spectroscopy. First, more than one Andreev bound state
(next to a prospected center Majorana state) is required to realize the
inter-state correlation defining the quasi-oscillatory spectral density profile, i.e., a
single-channel limit due to too strong transverse size quantization would
prevent our approach from working. The flipside of the same coin is that a large
spacing between individual Andreev levels implies their accurate resolvability.
In this regard, the nanowire platform appears to be superior to the TI vortex. 
Second, geometric constrictions may  prevent the placement of more than perhaps a single
electrode. However, compared to the TI vortex with its rigid flux quantization,
there now is larger freedom to apply  magnetic fields of different strength and
orientation. A combination of gate voltages and field strength parameters  may
go a long way in realizing a sizeable ensemble. All in all, we consider hybrid
nanowires as promising candidates for a statistical extension of the already
existing spectroscopic single-shot measurement setups.

\emph{Iron-based superconductors} \cite{Zhang2016,Wang2018,Machida2019,Zhu2020}
such as $\mathrm{FeSe}_{x}\mathrm{Te}_{1-x}$ 
define another Majorana platform of much current interest.  
They are expected to host topological surface states \cite{Zhang2016} where defect-induced 
vortices in a magnetic field bind Majorana states, similar to the TI vortex case.
(The short superconducting coherence length, \(\xi\approx (2-12)\,\)nm \cite{Petrovic2010,Wang2018}, here implies a much smaller vortex area.)   
Noting that the available samples show a large number of surface defects,
rather than generating the disorder ensemble by system parameter variations, 
the averaged spectral density $\langle \rho(\epsilon)\rangle$ could be
directly measured by performing tunneling spectroscopy on sufficiently large samples
containing many defects.  Conductance histograms 
of this type have already been published \cite{Machida2019,Zhu2020}, and 
with further refinements may allow to implement our approach. 

To summarize, several other currently investigated Majorana hardware
platforms appear to be suitable for the statistical spectroscopy approach. 
In either case, numerical simulations for realistic system parameters are relatively
straightforward to implement as trial runs prior to an experimental test. 

\bibliographystyle{aipnum4-1}
\bibliography{biblio}